\documentclass[a4paper, 12pt]{article}

\usepackage[english]{babel}
\usepackage[utf8x]{inputenc}
\usepackage[T1]{fontenc}

\usepackage[a4paper,top=3cm,bottom=3cm,left=3cm,right=3cm,marginparwidth=1.75cm]{geometry}

\usepackage{avant} 

\usepackage{color}
\usepackage{amsmath}
\usepackage{graphicx}
\usepackage[colorlinks=true, allcolors=blue]{hyperref}
\usepackage[page, title]{appendix}
\usepackage{graphicx}
\usepackage{gensymb}
\usepackage{xfrac}
\usepackage[font=small]{caption}
\usepackage{authblk}

\title{Diffusive external light-trap for solar cells} 
\author[1]{Ido Frenkel}
\author[1]{Shilpi Shital}
\author[1,*]{Avi Niv}
\affil[1]{\small{Department of Solar Energy and Environmental Physics, Jacob Blaustein Institutes for Desert Research, Ben-Gurion University of the Negev, Israel.\newline}}
\affil[*]{\small{Corresponding author: aviniv@bgu.ac.il}}
\begin{document}
\maketitle

\begin{abstract}
	\noindent The ability to absorb light is indispensable for high efficient solar power generation. This places conflicting requirements on the structure of a solar cell: On one hand it needs to have thick active layers to absorb more of the available sunlight while on the other it needs thinner ones for better charge transport. This dilemma stands in the way of any semiconductor from ever achieving its full potential as a solar cell material. Recently, external light-traps have emerged as a cost-effective solution for this dilemma by being able to decouple the optics from the electronic aspects of the power-generating process in the cell. In this paper we study the effectiveness of external light-traps with diffusive inner reflecting walls. A closed-form expression for the absorption of a cell, once placed inside or against the wall of such trap, are derived within the frame work of non-imaging and statistical ray optics. Results come to good agreement with the measured radiance from a modeled-trap. Our formulation indicates that a silicon cell, for example, may lose half of its optical absorption by becoming thinner, but will still regain its overall absorption if placed inside such an external trap, and thus gain higher $V_{oc}$, while maintaining $I_{sc}$ that would lead to higher efficiency. The effect of the trap for hydrogenated amorphous silicon and perovskite cells is also discussed with similar conclusions. The ability to decouple the optics from the electronics aspects of the cell raises the possibility of materials so far doomed obsolete may be introduced back on stage by using such trap. Being extraneous, the proposed trapping mechanism is, in fact, additive to whatever internal mechanisms the cell may already possess. We also use ray-tracing simulation to study more complex shaped traps, beyond the reach of our analytical formulation. Results indicate that given the correct design, external light-traps may present a genuine opportunity for high efficiency cost effective solar power production.
\end{abstract}
\newpage
\section{Introduction}
	Absorbing as much of the available sunlight as possible while efficiently transporting the resulting charge carriers to the external circuitry is the main challenge of high-efficiency solar cells. These are, however, conflicting requirements since substantial optical absorption requires thickness while charge transport is better at thinner layers. To relax this dilemma, light-trapping mechanisms have been devised which allow a relatively thin semiconductor to absorb as if it was a thicker one. Most successful in that regard are the ray randomization schemes that traps light within the cell's interior by total internal reflection (TIR) of rays that propagates at angles beyond some angular escape cone. The accumulative effect of this internal reflection, as first analyzed by Yablonovitch \cite{Yablonovitch1982}, causes the ray path-length inside the semiconducting slab to effectively approach $4n^2$ its thickness, where $n$ is the refractive index of the cell's material.
	
	\par Being an indirect semiconductor, silicon has a relatively small absorption coefficient of $\approx 1 cm^{-1}$ and an exceptionally large refractive index of $~3.6$, both given here at its bandgap wavelength. This combination makes silicon a perfect match for TIR based trapping since, in this case, $4n^2 \approx 50$. Having 50-fold path-length enhancement, means the cells can be made significantly thinner without sacrificing its absorption and hence its current. Due to the excellent charge transport abilities of silicon this translates to high efficiency. Combined with low cost of material, it is not surprising that silicon has the lion's share of the solar cell market.
	
	\par What about other materials though? Potentially cheaper or with smaller ecological footprint \cite{Mulvaney2014,dubey2013socio}. Among such materials are copper indium gallium selenide (CIGS) or cadmium tellurium (CdTe) of the so called "thin film" technology that scavenge whatever is left of the market. Other, more conceptual materials such as organic, oxide, or perovskite based technologies \cite{chen2018large,ruhle2012all,meng2018organic} - all typically have $n<3$ and in some cases even $n\approx2$. In addition, all have relatively high absorption coefficient - a combination that makes TIR  based trapping un-effective. One might argue that the large absorption coefficient might compensates for the inability to trap light and even, perhaps, make it redundant. This is, however, not the case since all of the above have lesser charge transport ability which forces them to be too thin for perfect light absorption. Indeed, in a recent study, Polman et al. reviewed record efficiency cells made from different materials, including the above mentioned ones \cite{Polman2016}. There, it was found that all present cell designs and materials, including the conceptual ones, have room for improvement in terms of both their optical and electronic aspects - a conclusion that encompasses also the state-of-art c-Si and GaAs cells. This highlights the bottleneck placed upon present solar cell technologies by the conflicting requirements of optics and charge-transport. It is important to note that other light-trapping schemes, not based on TIR, have also been devised. Notable in that regard are those based on plasmonic or photonic scattering \cite{Polman2008,Polman2012a,collin2018new}, or emission restriction \cite{Munday2014,Niv2012}. None, however, have yet been adopted commercially due their costly fabrication and the potential burden they present to the already intricate interfaces of the solar cell material slab.
	
	\par Recently, a relatively old approach has been revived: The external light-trap. Here, the entire cell is placed within a device that redirects whatever is not absorbed at first instance back to the cell. Renew interest comes from the fact that external light-traps circumvents the above mentioned dilemma by decoupling the optics from charge transport mechanism. After being first presented in the early 1990s, \cite{luque1991confinement}, it re-emerged as means of using the light of radiative recombinations, a process known as photon recycling \cite{Braun2013}, but was soon harnessed to recycle a far greater light source, which is the cell's reflection \cite{Weinstein2015, Niv2016a}. Soon after, its full potential as a mean of decoupling optics from charge transport was acknowledged \cite{Dijk2016}; Unlike TIR plasmonic or photonic light-trapping schemes, this one does not sacrifice the simplicity of cell's layout in favor of better absorption. On the contrary, here the cell is liberated to achieve its optimal charge separation and transport abilities on the expense of its optical absorption. The unabsorbed light is not lost, however, but is redirected toward the cell by this external light-trapping device. Two main approaches for realizing external traps have emerged: The first is based on specular reflection from the inner light-trap walls \cite{luque1991confinement,Braun2013}. This one benefits from the high reflection of polished metalize surfaces but requires precision manufacturing and careful placement of the cell \cite{Weinstein2014a, Dijk2016}. The second is based on diffused back-scattering inner walls \cite{Niv2016a}. While being a bit less reflective, this one is more tolerant to the trap geometry and to the exact placement of the cell within it.
	
	\par Surprising as it may sound, comprehensive treatment of external light-traps has yet to emerge. At the time of their initial presentation they were only analyzed in broad terms but not at the level of a full performance characterization \cite{luque1991confinement}, while later treatments used ray-optics simulations to study specific configurations \cite{Weinstein2015, Dijk2016}. Here we aim to use the tools of non-imaging and statistical-ray optics to derive a closed-form expressions for the absorption of a cell that is placed inside a diffusive light-trap shaped as arbitrary as possible. To do so, we use the single flux-balancing approach, first used by Yablonovitch to derive the $4n^2$ effective ray-path enhancement factor \cite{Yablonovitch1982}. The major advantage of this approach is being least respective of the device geometry and therefore more general. Alternatives like the two flux or radiation-transport methods perhaps take better account of losses but are based on having well defined propagation directions and, therefore, are more fit to a slab \cite{Gee1988,Green2002,dahan2013}. For an enclosure, such as the external trap, there is no preferred direction of light rays so it is not clear how to implement these methods.
	
	\par In the following we first find the radiance $L (W sr^{-1} m^{-2})$ of the light inside an empty enclosure and then use it to find the absorption of a cell that is placed inside such a device. Two cases would be considered: wall-mounted and center-mounted cell. Our analysis indicates that the center mounted configuration has the upper hand in terms of absorption enhancement. We proceed by matching our results against measured data from different configurations of an external trap. The good agreement that emerges confirms the validity of our derivations. Finally, we use ray-optics simulation to study more realistic configurations of an external light-trap, other than those tested upon. The results of which validate our analytic approach in some sens while pointing to its limitations just as well. Our findings indicate that a properly designed diffusive external light-trap may present a genuine opportunity for highly-efficient cost-effective solar power conversion schemes. 
	
\section{Non-imaging optics of empty diffusive external light-traps}\label{sec:empty_trap}
\begin{figure}
		\centering
		\includegraphics[width=0.8\textwidth]{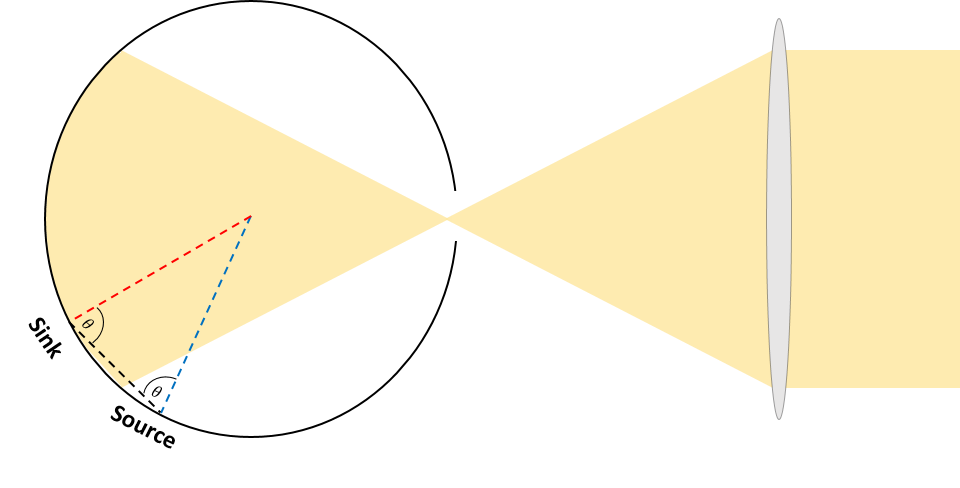}
		\caption{(A) An empty light-trap that is fed with focused radiation from some concentrator device. The isosceles triangle that is formed by a source, sink and center of the spherical light-trap is also depicted.}
		\label{fig:empty-cav-iso-triangle}
	\end{figure}

	External light-trap is an enclosure with highly reflecting inner walls and with a small port allowing light to enter and, unavoidably, also for some to escape, an example of which is shown in Fig. \ref{fig:empty-cav-iso-triangle}. The trap's inner walls may be specular-reflecting as polished metal or be diffusive-reflecting as white paint does. Each has its own advantages and downsides: The specular one typically has better reflection but requires precise manufacturing and careful alignment to the cell \cite{Weinstein2015, VanDijk2016a}. The diffusive kind may be, but not necessarily, less reflective but in turn is more tolerant to aberrations in its form. Also, due to the presence of surface roughening or front-contacts, the reflection from a solar cell may contain a significant diffusive component. In that case, randomized ray-direction are expected to emerge within the trap even if the cell is placed inside a specular inner wall reflecting device. In that sense, analysis based on the assumption of randomized ray direction is more realistic. Lastly, assuming that rays have randomized directions allow us to forsake the tracing of an individual ray in favor of fluxes of given power and angular distribution. This major simplification is the key to our ability to derive an analytical expressions of the trap's optical performance. 

	\par First let us consider an empty trap with diffused-reflecting inner walls, a schematic depiction of which is shown in Fig. \ref{fig:empty-cav-iso-triangle}: Light with a radiant flux power $P_{in} (W)$ enter the trap by being focused onto its port. The precise means by which it is focused are irrelevant - the depicted lens is for illustrative purpose only. Once inside, light propagate until it reached the wall opposite to the port. Upon impinging radiant flux equal to $\alpha_{w}P_{in}$ is absorbed by the wall, where $\alpha_{w}$ is the normalized wall absorption. The rest, which is $(1-\alpha_{w})P_{in}$, is back scattered into the trap. Due to this back reflection, the trap is "filled" with isotropic (diffused) light. Being a passive device, all that enters must leave; the available depletion routs in this case are wall absorption or escape through the port. The amount of radiant flux that is absorbed by the wall is:
		$$\alpha_{w}LF_{w}A_{W}$$
	and the radiant flux that leaves through the port is:
		$$LF_{p}A_{p}.$$
	Here, $L$ denotes the \emph{radiance} $(W sr^{-1} m^{-2})$ of the diffused light filling the trap enclosure, and $F_{w,p}$ and $A_{w,p}$ are the respective wall or port \emph{view-factor} $(sr)$ and \emph{area} $(m^2)$. The view-factor and area are where the trap geometry enters our formalism. 
	
	\par Let us now consider the back scattered light after first wall impingement as the \emph{source} of \emph{diffused} radiance filling the trap. In that case, the following flux-balance should holds:
	\begin{equation*}
		(1-\alpha_{w})P_{in}=\alpha_{w}LF_{w}A_{w}+LF_{p}A_{p}.
	\end{equation*}
		The radiance of the diffused light that files the trap's empty enclosure is thus:
	\begin{equation}\label{eq:L-empty}
		L=\frac{(1-\alpha_{w})}{\alpha_{w}F_{w}A_{w}+F_{p}A_{p}}P_{in}.
	\end{equation}
	Equation (\ref{eq:L-empty}) has the generic form of the flux-balance approach but includes the view-factor which is inherent from non-imaging optics. Having a well-defined light-trap means knowing the view-factors, areas, and wall absorption whereupon Eq. (\ref{eq:L-empty}) determines the radiance inside the light-trap. 
	
	\par In order to proceed, we must have closed-form expressions for the view-factors of the wall and port. For that matter, let us consider a ray impinging the wall or the port at some angle $\theta_{sink}$. The ability of this ray to be absorbed by the wall or be transmitted through the port is proportional to $\cos(\theta_{sink})$. Likewise, let us assume a ray scattered off the wall at an angle $\theta_{source}$. The radiance that this ray carries (within an infinitesimal solid angle cone) is proportional to $\cos(\theta_{source})$. All together then, the differential view-factor of an infinitesimal solid-angle and unit-area is:
	\begin{equation}\label{eq:dV-factor}
	dF=2\pi\sin(\theta_{sink})\cos(\theta_{sink})\cos(\theta_{source})d\theta_{sink},
	\end{equation}
	where azimuthal symmetry has been assumed. If the trap geometry is known then $\theta_{source}$ can, at least in principle, be expressed as a function of $\theta_{sink}$. Integration over $\theta_{sink}$, in this case, gives the view-factor at the sink point.
	
	\par Finding $\theta_{source}$ and a function of $\theta_{sink}$ can be a tedious task except when symmetry considerations may be invoked. One such case, perhaps the simplest of them all, is that a spherical shaped trap, as in Fig. \ref{fig:empty-cav-iso-triangle}. For this particular form, the ray leaving the source towards the sink and the normal to the sink and source points forms an isosceles triangle whereupon $\theta_{sink}=\theta_{source}$. Integrating Eq. (\ref{eq:dV-factor}) over $\theta_{sink}$, in this case, readily gives:
	\begin{equation}\label{eq:V-factor-empty}
		F=\frac{2\pi}{3}.
	\end{equation}
	Plugging this back to Eq. (\ref{eq:L-empty}) while considering that in this case $F_{w}=F_{p}$, the radiance inside a spherical trap is found to be: 
	\begin{equation}\label{eq:L-empty-sphere}
		L=\frac{3}{2\pi}\frac{(1-\alpha_{w})}{\alpha_{w}A_{w}+A_{p}}P_{in}.
	\end{equation}
	Having a closed-form expression for the radiance allows us to calculate the power coupled to each of the above mentioned depletion mechanisms; the wall or the port. Let us, for example look at the power exiting port:
	\begin{equation*}
		P_{port}=F\,A_{p}\,L=\frac{A_{p}(1-\alpha_w)}{A_p+\alpha_{w}A_{w}}\,P_{in}.
	\end{equation*}
	It is seen that if the wall is non-absorbing this becomes: $P_{port}=P_{in}$ - as it indeed should. In the following we shall use similar considerations to find the radiance flux the cell absorb by being placed inside a spherical light-trap.

\subsection{Spherical diffusive light-trap with a wall-mounted cell}
\begin{figure}
		\centering
		\includegraphics[width=0.8\textwidth]{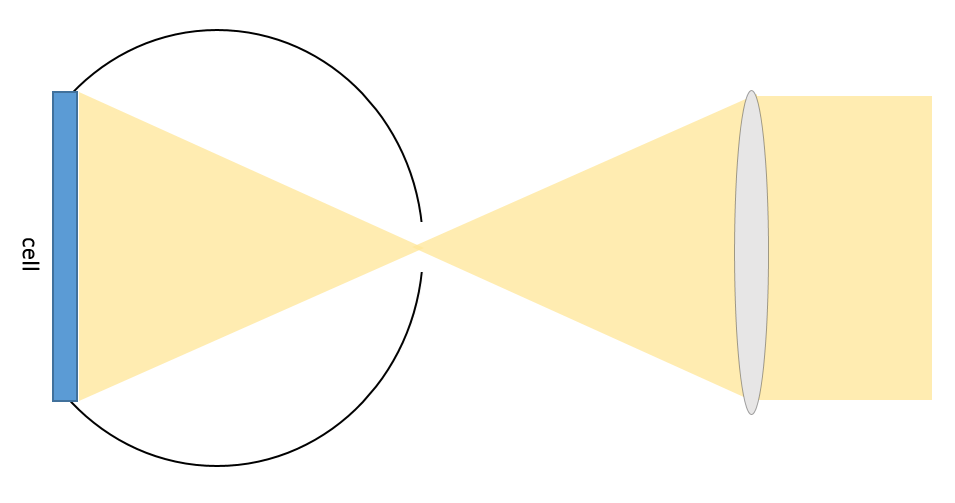}
		\caption{Schematic view of a reflector with a wall-mounted cell. The cell may receive fraction of the light directly while absorbing also diffused light-trapped inside the external light-trap.}
		\label{fig:wall-mounted-cell}
	\end{figure}

	Let us examine the more practical case of a spherical diffusive light-trap with a cell mounted on its wall, as depicted in Fig. \ref{fig:wall-mounted-cell} - A configuration that appears in two of the three previous treatment of external traps that we have been able to find \cite{luque1991confinement,Dijk2016}. When analyzing this case, we will consider the cell to be posited opposite to the port define a dimensionless parameter $\beta=A_{c}/A_{beam}$, where $A_{beam}$ and $A_{c}$ are the area illuminated by the beam and that of the cell, respectively. This way $\beta P_{in}$ is the fraction of incoming radiance flux that falls upon the cell and, accordingly:
	\begin{equation*}
		\beta \alpha_{c} P_{in},
	\end{equation*}
	is the power absorbed by the cell at first incidence. The back-scattering from the cell then would be:
	\begin{equation*}
		\beta (1-\alpha_{c}) P_{in},
	\end{equation*}
	where $\alpha_c$ is the normalized cell absorption. If the cell does not cover the entire illumination spot, such that $\beta<1$, then radiance flux of $(1-\beta)(1-\alpha_w)P_{in}$ is back-reflected form the exposed trap wall. The total diffused radiance flux that enters the trap is thus:
	\begin{equation*}
		[\beta (1-\alpha_{c})+(1-\beta)(1-\alpha_w)] P_{in}=[(1-\alpha_w)+(\alpha_w-\alpha_c)\beta]P_{in},
	\end{equation*}
	The depletion routs available for this flux are the wall and port, as before, but also the cell absorption, which in turn absorbs $L\,F_{c}\,A_{c}$ from the randomized radiance flux that fills the light-trap. Balancing the incoming and outgoing fluxes, considering that the cell's view-factor is identical to that of the port and wall, gives the radiance that is established in the trap in this case:
	\begin{equation*}
		L=\frac{3}{2\pi}\,\frac{(1-\alpha_w)+(\alpha_w-\alpha_c)\beta}{A_p+\alpha_{w}A_{w}+\alpha_{c}A_{c}}P_{in},
	\end{equation*}
	and the corresponding cell absorption is then: 
	\begin{equation*}
		\frac{\alpha_{c}A_{c}[(1-\alpha_w)+(\alpha_w-\alpha_c)\beta]}{A_p+\alpha_{w}A_{w}+\alpha_{c}A_{c}}P_{in}.
	\end{equation*}
	The total cell absorption is the above term added with the flux absorbed at first instance:
	\begin{equation}\label{eq:wall-mounted-cell-abs}
		P_{c}=\left \{\alpha_{c}\beta+\frac{\alpha_{c}A_{c}[(1-\alpha_w)+(\alpha_w-\alpha_c)\beta]}{A_p+\alpha_{w}A_{w}+\alpha_{c}A_{c}} \right \}P_{in}.
	\end{equation}
	The above equation gives the power absorbed by a cell which is placed on the wall of a spherical diffusive trap. This radiance is characterized by the cell ($\alpha_{c}$,$A_{c}$) and trap ($\alpha_{w}$,$A_{w}$,$A_{p}$) parameters as well as the quality of illumination ($\beta$). It is easy to see that the flux absorbed by the cell equals the incoming one ($P_{c}=P_{in}$) for a perfectly absorbing cell ($\alpha_{c}=1$) that is completely covered by the illuminated spot ($\beta=1$), as it should. More interesting, perhaps, is a non perfectly absorbing cell mounted inside a trap with perfectly-reflecting inner wall ($\alpha_{w}=0$). The cell absorption in this case is:
	\begin{equation*}
		P_{c}=\left \{\alpha_{c}\beta+\frac{\alpha_{c}A_{c}[1-\alpha_c\beta]}{A_p+\alpha_{c}A_{c}} \right \}P_{in} \quad (\alpha_{w}=0).
	\end{equation*}
	If, in addition, the port is so small such that its effect can be neglected, we end up with:
	\begin{equation*}
		P_{c}=\left \{\alpha_{c}\beta+1-\alpha_c\beta \right \}P_{in}=P_{in} \quad (\alpha_{w}= A_{p}=0).
	\end{equation*}
	This points to one of the major advantages of external light-traps: The cell will take most of the radiance as long as its absorption is larger then the combined wall and port one, no matter how small it is.
	
	\par
\begin{figure}
		\centering
		\includegraphics[width=.8\textwidth]{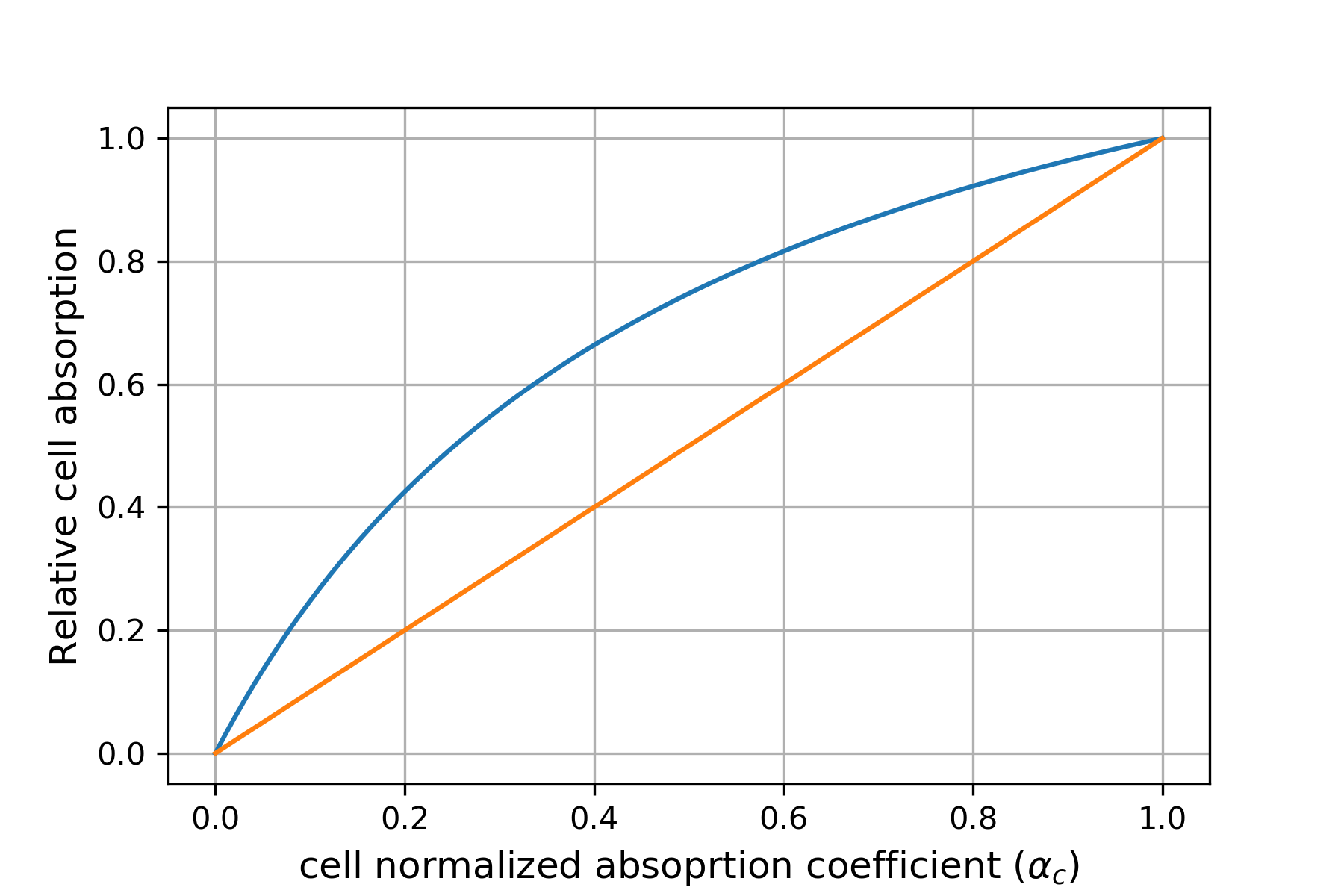}
		\caption{The relative absorption ($P_{c}/P_{in}$) of a wall mounted cell (blue) and a directly illuminated cell without a reflector (orange) for the following parameter set: $\alpha_{w}=0.05; \beta=1; A_w=100; A_c=1; A_p=0.01.$}
		\label{fig:wall-mounted-cell-absorption}
	\end{figure}

 To demonstrate this property of the trap we plot in Fig. \ref{fig:wall-mounted-cell-absorption} the relative absorptions of a cell placed inside a trap to that of a directly exposed cell as a function of cell absorption $\alpha_{c}$. The trap aided absorption follows from  to Eq. (\ref{eq:wall-mounted-cell-abs}), while the exposed cell absorption is given by $\alpha_{c}P_{in}$ in this case.
	Calculations were carried for: $\alpha_{w}=0.05;\ \beta=1;\ A_w=100;\ A_c=1;\ A_p=0.01$. The external trap ability to boost the native cell absorption is clearly observed. It can be seen that a 60\% absorbing cell would be boosted to more than 80\% by the trap. Even a cell with 90\% absorptivity would benefit somewhat from the trap. This property of the trap enables consideration of cells that excels in charge separation and transport on the expanse of its their optical absorption. Note that it doesn't matter, from the trap point of view, weather absorption was reduced due to thinner active layer or due to a denser front contact grid.

\subsection{Diffusive light-trap with a center mounted cell}
\begin{figure}
		\centering
		\includegraphics[width=.8\textwidth]{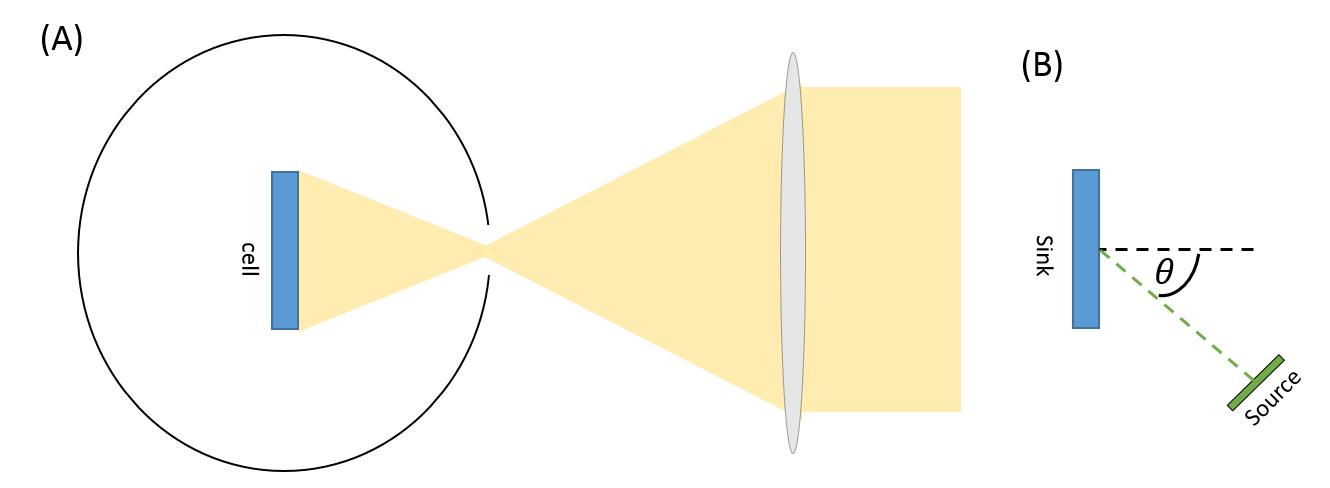}
		\caption{An illustration of a reflector with a center-mounted cell.}
		\label{fig:center-mounted-cell}
	\end{figure}

	Next, let us study a center-mounted cell as shown in Fig. \ref{fig:center-mounted-cell}(A). This case is identical to the wall-mounted case in every aspect but the cell's view-factor at its new location. Central mounting inside a spherical enclosure means that, to a good approximation, the cell receives radiation emitted always at right angle from the source points surrounding it, as shown in Fig. \ref{fig:center-mounted-cell}(B). This removes the source-related projection factor, such that the differential view-factor is given now by:  
	\begin{equation*}
		dF_{center}=2\pi\sin(\theta_{sink})\cos(\theta_{sink})d\theta_{sink},
	\end{equation*}
	 Carrying the $0-\pi/2$ integration over $\theta_{sinc}$ gives the more common form af the view-factor:
	\begin{equation*}
		F_{center}=\pi A_{c}.
	\end{equation*}
	Note that this view-factor is 1.5 times the view-factor of a wall-mounted cell. Balancing the incoming and exiting fluxes in this case gives the radiance that fills the diffusive spherical trap when the cell is mounted at the center of the spherical trap:
	\begin{equation*}
		L_{center}=\frac{3}{2\pi}\,\frac{(1-\alpha_w)+(\alpha_w-\alpha_c)\beta}{A_p+\alpha_{w}A_{w}+\frac{3}{2}\alpha_{c}A_{c}}P_{in},
	\end{equation*}
	The radiance that fills the light-trap in this case is, thus, a tad smaller relative to the wall-mounted case due to the larger denominator of this case. However, what we are looking for is the cell's absorption in this case:
	\begin{equation}\label{eq:center-mounted-cell-abs}
		P_{c}=\left \{\alpha_{c}\beta+\frac{3}{2}\, \frac{\alpha_{c}A_{c}[(1-\alpha_w)+(\alpha_w-\alpha_c)\beta]}{A_p+\alpha_{w}A_{w}+\frac{3}{2}\alpha_{c}A_{c}} \right \}P_{in}.
	\end{equation}
	Also here the first factor within the curly brackets is the absorption of direct illumination and the second is the contribution from the diffused radiance filling the trap.	It is seen that center-mounted cell's view-factor changes the absorption with respect to the wall-mounted one from Eq. (\ref{eq:wall-mounted-cell-abs}) in two ways: It makes the absorption of diffused radiance $3/2$ times larger since sources on the trap wall are always at right angle to the cell but also makes a larger denominator for the same reason.
	In other words, better view factor increases the cell absorption that, in turn, reduces the radiant flux inside the cavity, but even with this reduced radiant flux, the cell absorption is greater for center-mounted cell.
	Due to the presence of other depletion mechanisms in the denominator, the overall effect of the center-mounting view-factor is to enhance the absorption.
	\par
\begin{figure}
		\centering
		\includegraphics[width=.8\textwidth]{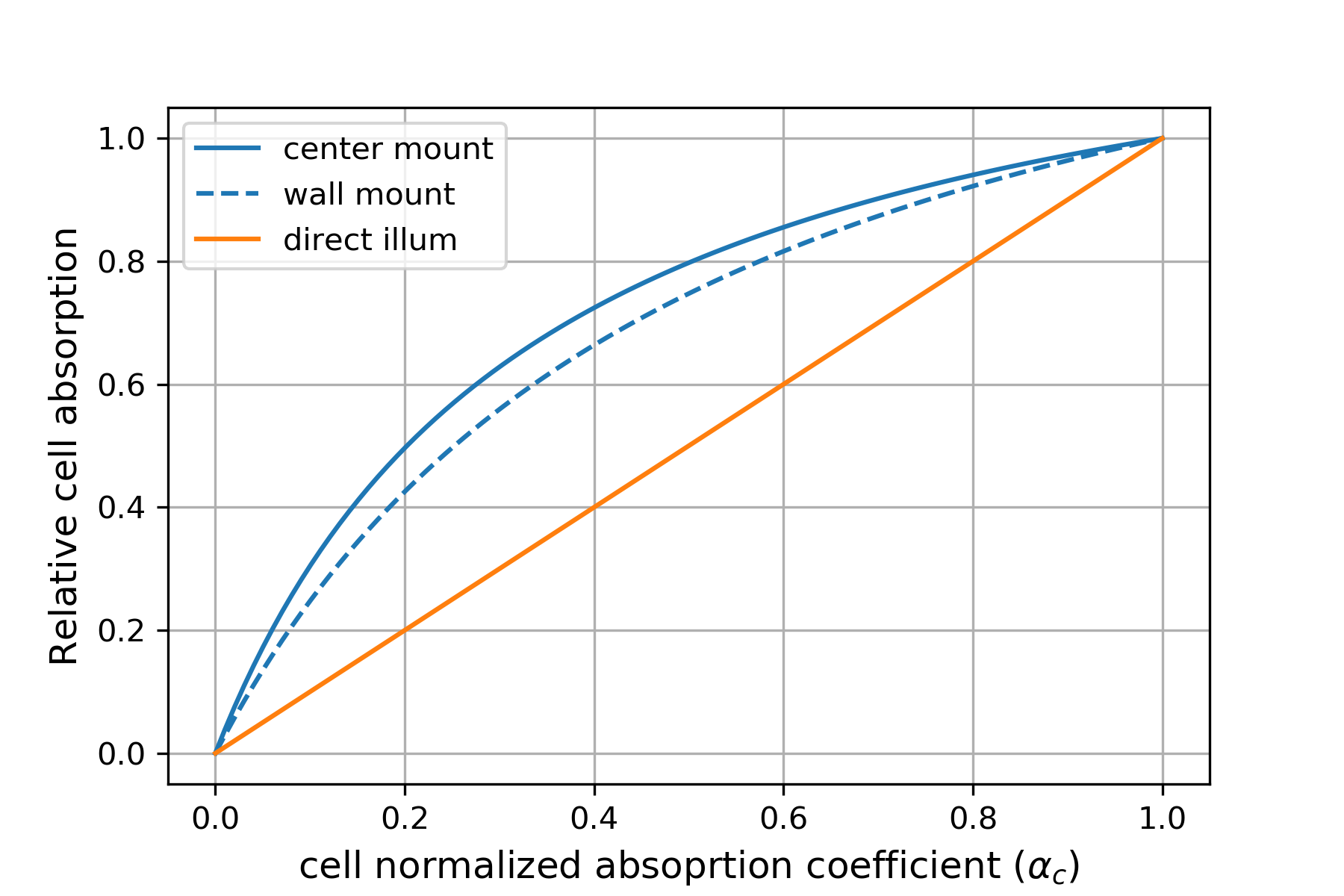}
		\caption{The relative absorption ($P_{c}/P_{in}$) of a center mounted cell (blue) and a directly illuminated cell without a reflector (orange). For comparison also the case of a wall-mounted cell is shown in dashed blue. The following set of parameters was chosen: $\alpha_{w}=0.05; \beta=1; A_w=100; A_c=1; A_p=0.01.$}
		\label{fig:center-mounted-cell-absorption}
	\end{figure}

 To show this effect we compare in Fig.\ref{fig:center-mounted-cell-absorption} the relative absorption of a center-mounted cell to a direct illuminated and a wall-mounted one, all as a function of the cell's absorption $\alpha_{c}$. The system parameters, in this case, are identical to those used for the plot in Fig. \ref{fig:wall-mounted-cell-absorption}. The advantage of the center-mounting configuration clearly emerges: A cell with 60\% bare absorptivity, for example, absorbs about 85\% if mounted in the center of the trap relative to little over 80\% of the wall-mounted one. This indicates that the hemispherical traps proposed in \cite{luque1991confinement} might be the best option since area is minimized and central view-factor is obtained but for a wall mounted cell, which is more desirable for practical reasons.

\section{Experimental results} \label{sec:exp res}
\begin{figure}[h]
		\centering
		\includegraphics[width=.8\textwidth]{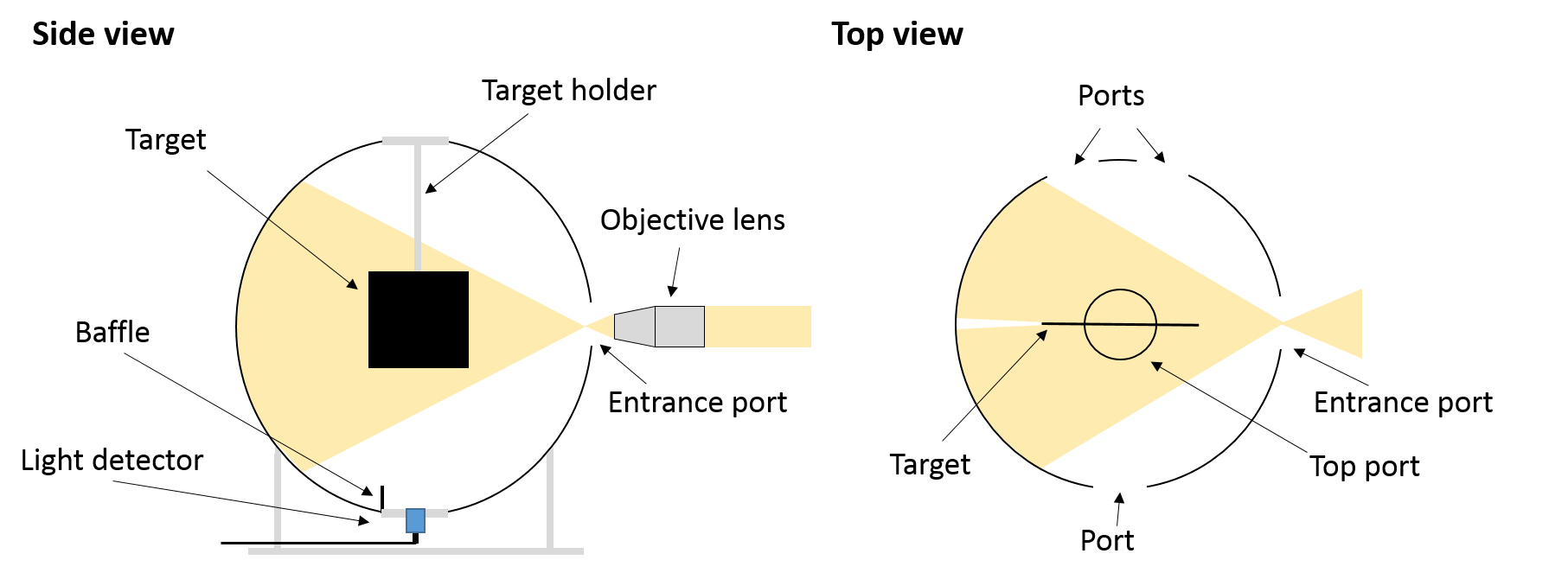}
		\caption{Schematic depiction of the testing apparatus.}
		\label{fig:setup}
	\end{figure}

\begin{figure}[h]
		\centering
		\includegraphics[width=.8\textwidth]{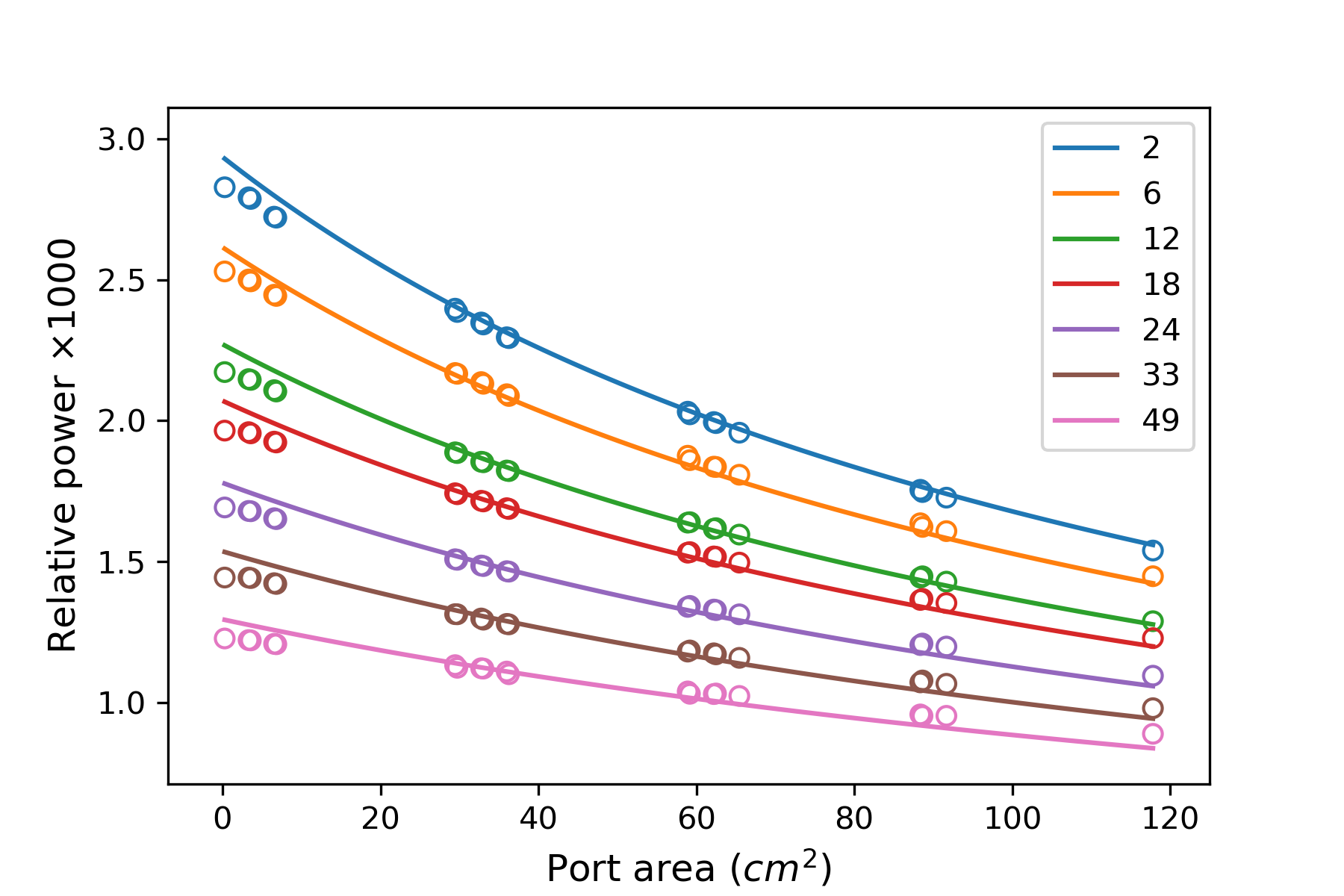}
		\caption{Experimental (circles) and calculated (lines) results versus port-area. The notation of \emph{Relative power} refers to the ratio of wall-mounted to incoming power, for the measured and calculated data alike. Colors indicate the different center-mounted absorber area in $cm^2$. The fitting included an additional holder absorption and direct illumination factors ranging from $0$ for the $2cm^2$ cell and up to $30\%$ for the $49cm^2$ foils}
		\label{fig:theory-exp}
	\end{figure}

	To test our formulas, we have utilized the setting depicted in Fig. \ref{fig:setup}: A 14.5 cm diameter spectroscopic integrating sphere (UPB-150-ARTA from Gigahertz-Optik) was used as a modular external light-trap. The effect of a center mounted cell was mimicked by placing thin absorbing foils on the sample holder of the sphere. These foils were mounted so that their surface was parallel to the illumination direction to minimize the direct-illumination factor $\beta$. This way the inner trap radiance was most affected by the wall and port. Changing the foil area represented different cell absorption. The sphere had 3 modular ports that could be fitted with different port-covers of various hole diameter, including blank ones. Overall, 72 combinations of port-cover emerged representing 47 different port-areas. Therefore, inner-trap radiance was measured for different foil area with the port-area as the control parameter. Light from a halogen lamp was focused onto one of the ports. Accordingly, this port was never completely blocked neither was the beam stopped by the any port-covers that were fitted there. Power was registered by a wall-mounted detector and at the entrance port. This way, information about the power confinement abilities of the integrating sphere, now acting as a trap, were gathered.
	Fig. \ref{fig:theory-exp} shows in circles the registered wall to incoming power ratio as a function of port-area. Solid lines shows the predicted relative power according to our formalism, that was modified to account for the absorption of a wall-mounted detector given a center mounted absorber. More details of the absorption formula for this case and the fitting parameters are given in appendix \ref{app: model and fitting}. The overall good agreement confirms our formulation of the optical behavior of a diffusive external light-trap.

\section{Discussion}
	The good agreement between the prediction of our formalism and measured data encourage us to study what could be the benefit of using an external light-trap with a solar cells made from different materials such as silicon (Si), thin-film hydrogenated amorphous-silicon (a-Si:H), and finally the emerging perovskite. To do so, we used an analytical cell model based on Shockley and Queisser detailed balance approach \cite{Queisser1961}.
	%
	The model uses the absorption of that cell to estimate the open circuit voltage $V_{oc}$, short-circuit current $I_{sc}$, and therefore the power and efficiency. The model does so for a given bandgap, non-radiative recombination rate (of the Shockley-Read-Hall type), thickness, and internal light-trapping mechanism that is used for the cell. The source of radiant flux was a $6000K$ black-body and four kinds of light-trapping were considered: Two are internal in the form of a perfect back-reflector and TIR based trapping and two are the same back-reflection and TIR equipped cells but now also with and an external trap. Details of the analytic cell modeling and the implementations of different light-trapping schemes are given discussed Appendix \ref{app:cell-model}.
	\par
\begin{figure}[h]
		\centering
		\includegraphics[width=.6\textwidth]{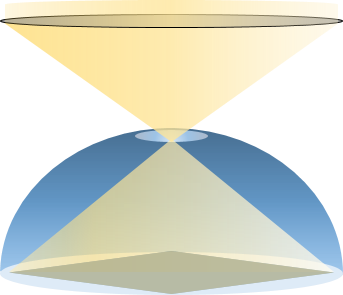}
		\caption{Illustration of the hemispherical external light-trap and the square cell inscribed by the circular its base.}
		\label{fig:dome}
	\end{figure}

\begin{figure}[h]
		\centering
		\includegraphics[width=1\textwidth]{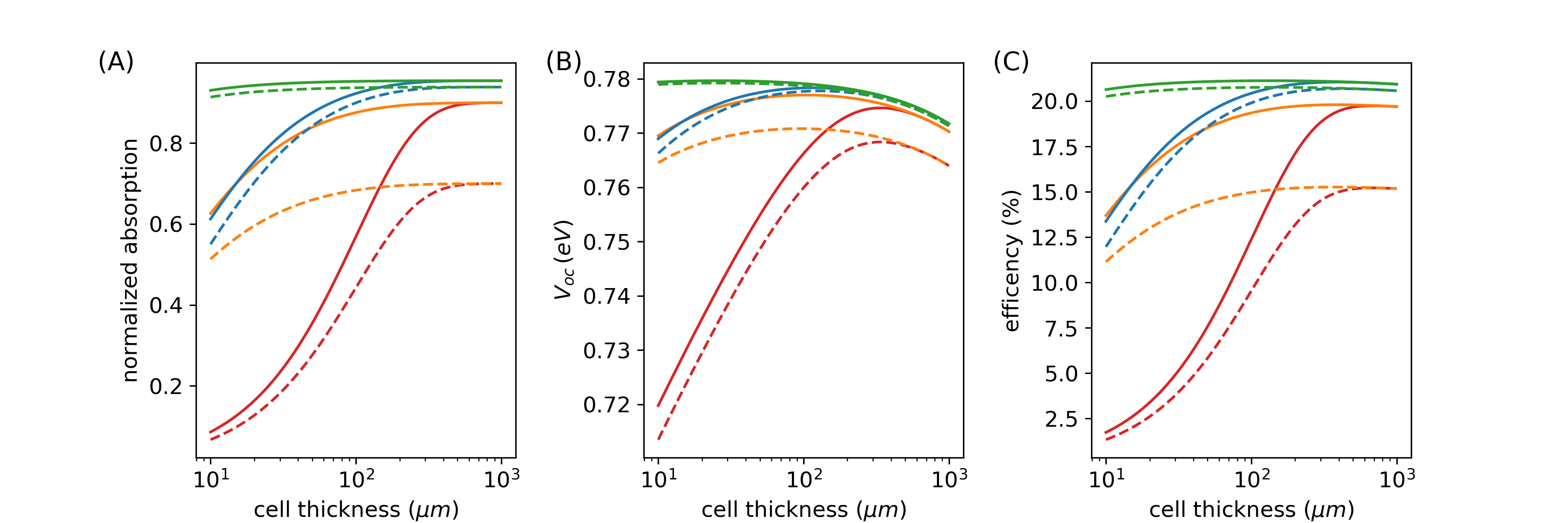}
		\caption{(A) Normalized absorption, (B) $V_{oc}$, and (C) efficiency versus crystalline silicon thickness. Different light-trapping approaches are coded in color: Only back reflector in red, Lambertian in orange, external trapping in blue, and finally a combined Lambertian and external trapping approach in green. Solid or dashed lines represent the result of having $10\%$ or $30\%$ front surface reflection, respectively.}
		\label{fig:si-cell}
	\end{figure}

 To facilitate a meaningful comparison between the different trapping methods, the internal back-reflection of the TIR-based approach and the wall reflection of the external trap were assigned an identical value of $R_{bsr}=1-\alpha_{w}=0.98$. With this in mind, it soon became clear that the critical design requirement is minimizing the trap wall area since $\alpha_{w}A_{w}$ is the dominant loss mechanism. We, therefore, considered a hemispherical light-trap, as shown in Fig. \ref{fig:dome}, which has small area and the better view-factor of the center-mounted cell. From a practical point of view, this trap also has the advantage of simple cell mounting as for the wall mounted case. The cell area was taken to be the larger square inscribed by the circular base of the hemisphere, which gave a cell to wall area ratio of $\approx 1:3.712$. Rectangular cross-sectioned illumination was assumed such that $\beta=1$. Also, cell to port-area was kept at fixed 100:1 ratio. Finally, $4\%$ front surface reflection was considered for the external trap due to possible imperfect reflection from the unspecified concentrator.
	\paragraph{Silicon cell:} First, let us examine the performance of a silicon cell that is equipped with either internal or external trapping schemes. Accordingly, the following parameters have been fed to our cell-model: bandgap  $E_g=1.1 eV$, average absorption $\alpha=50\,(cm^{-1})$, refractive-index $n=3.6$, and non-radiative recombination rate coefficient $4\times10^{-4}\,(cm^{-1})$. Auger non-radiative recombinations were not considered. In addition, a value of $R_{ext}=0.1$ was assigned for the front surface reflection of the cell expressing non-ideal anti-reflective coating and the possible effect of metal grid electrodes. This choice calibrate our model to reproduce the behavior of a good c-Si solar cell. Figure \ref{fig:si-cell} (A) shows the calculated absorption of the different trapping mechanisms that were considered: The two internal ones being a perfect back-reflector and TIR trapping while the two external ones are the former two but considering the respective cells to be placed inside the above mentioned hemispherical trap. As for the internal trapping schemes, the effectiveness of TIR based approach, shown in solid orange line, is clearly observed  - Full absorption is practically reached by $200 \mu m$, while as much as $1 (mm)$ is needed for the back-reflector case shown in red.
	\par The combination of perfect back reflector with the external-trap is shown in solid blue: It is higher then the TIR-based approach from $\approx20\mu m$ onward but reach its full advantage by $\approx 100 \mu m$. External-trap is thus able to make simple double-pass cell absorb more than a TIR-equipped one. The advantage comes, in this case, predominantly due to the external-trap ability to recycle the cell's front surface reflection, what the TIR based trapping cannot do.
	\par The effect of a combined TIR and external-trapping is shown in green solid line: Absorption is significantly better in this case. Here, practically full absorption is reached already by $10 \mu m$, demonstrating the additive quality of the external-trap.
	\par We have also repeated our calculations with much larger front surface reflection of $30\%$ instead of $10\%$; the results of which are shown with dashed lines while maintaining the same color-coding. It is seen that the external-trapping cases, shown in blue and green, are much less susceptible to front surface reflection. This demonstrate the ability to trade optical absorption for better charge transport and then to recover the lost absorption with an external-trap.
	\par We continue with the open circuit voltage, $V_{oc}$, of the above mentioned cases, as shown in Fig. \ref{fig:si-cell}(B). Here, once more, the TIR is better then the back reflector and comparable to the back-reflector combined with an external trap. At first $V_{oc}$ raises as the cell thins due to the reduction of nonradiative recombinations. After reaching its peak at $\approx 100 \mu m$, $V_{oc}$ eventually drops since the effect of vanishing absorption overcomes the reduction in bulk nonradiative recombinations. As for the combined TIR/external-trap case, due its superior absorption, not only it has the highest $V_{oc}$ but is also able to maintain it for silicon layer that is $10 \mu m$ thin. Also here, having larger front surface reflection, as shown with dashed lines, has a devastating effect on the internal TIR-based and back-reflector light-trapping but has negligible effect when external-trap is considered.
	\par The predicted efficiency, calculated as $FF(V_{oc})V_{oc}I_{sc}/P_{sun}$ for $P_{sun}=100(mW/cm^2)$ ,is shown in Fig. \ref{fig:si-cell}(C). Since $I_{sc}$ is proportional to the absorption and $FF$, the fill-factor, is only a weak function of $V_{oc}$, the efficiency trends like the absorption-$V_{oc}$ product. Accordingly, also here the external-trap is able to sustain absorption, $V_{oc}$, and thus the efficiency for low-absorption cell. The overall effect of the external-trap on a c-Si cell is that efficiency raises from $19.8\%$ to $21.1\%$ at Si layer thickness that reduces from $379\,nm$ to about $125\,nm$. Since the price of silicon is low, external-traps do not represent a significant improvement for this solar cell technology. It is remarkable, however, that the efficiency of a combined TIR and external trap is only a weak function of thickness, as seen by the green lines of panel (C), such that the cell maintains above $20\%$ efficiency down to Si thickness of $10\mu m$ - a significant reduction in material. One should also be mindful that the present analysis does not consider the effect of having larger front grid electrodes on the serial resistance or an Auger-type nonradiative recombinations on $V_{oc}$ and is, therefore, an underestimate of the possible effect of an external-trap.
	%
	\paragraph{Hydrogenated amorphous silicon:}
	Hydrogenated amorphous silicon, a-Si:H, is an important representative of the thin-film solar cell technology. Despite having a larger-then-optimum bandgap of about $1.7 eV$, it was considered an alternative to c-Si due to its simple large-area deposition methods. This technology was largely abandoned eventually, partially because of a light induced degradation mechanism - the Staebler-Wronski effect \cite{staebler1977reversible}. It was, however, acknowledged that this effect is less severe in thinner layers, be it due to formation of larger grains or due to the large electric-field across a thinner junction \cite{shah2004thin}. It would be, therefore, interesting to study what could be the effect of external light-trap on this kind of cell.
	\par
\begin{figure}[h]
		\centering
		\includegraphics[width=1\textwidth]{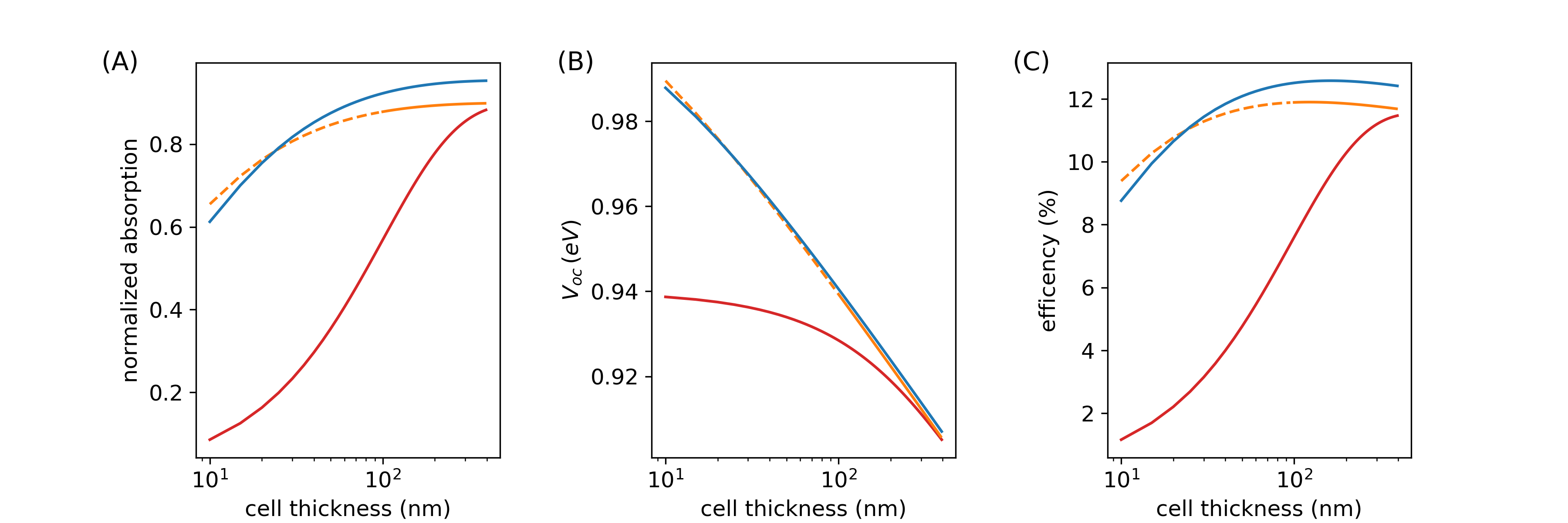}
		\caption{(A) Normalized absorption, (B) $V_{oc}$, and (C) efficiency versus a-Si:H layer thickness. Light trapping approach is color coded: Back reflector in red, Lambertian in orange, and external trapping in blue. The dashed orange line indicate layer thickness where Lambertian trapping looses its effectiveness due to invalidity of the ray-optics description of light.}
		\label{fig:a-si-cell}
	\end{figure}

 Figure \ref{fig:a-si-cell} shows the absorption, $V_{oc}$, and efficiency that are obtained by tuning our cell model to material parameters of a-Si:H, namely: $1.7 eV$ bandgap, $1\times10^{12}$ (relative) non-radiative recombination rate, $5\times10^{4} cm^{-1}$ absorption coefficient, and $n=4$ refractive index \cite{beaucarne2007silicon,palik1998handbook}. All other environmental and trap parameters are the same as those used with c-Si.
	\par
\begin{figure}[h]
		\centering
		\includegraphics[width=1\textwidth]{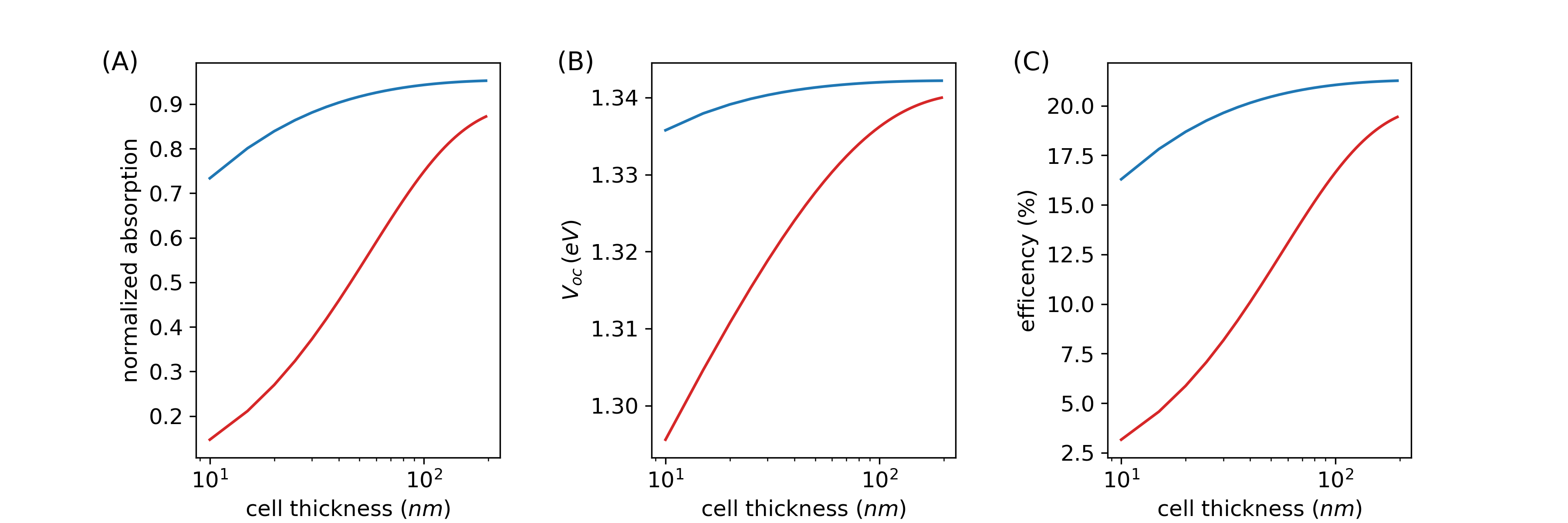}
		\caption{(A) Normalized absorption, (B) $V_{oc}$, and (C) efficiency versus a typical perovskite layer thickness. The effect of using a back-reflector is shown in red while that obtained from external trapping in blue.}
		\label{fig:perovskite-cell}
	\end{figure}

 Due to the large native absorption of a-Si:H, all light is seen to be absorbed with a $200 nm$ layer with a back-reflector. This reduces to less than a $100 nm$ if TIR or External light-trap are used (orange and blue, respectively). Also here, the clear advantage of external light-trap is its ability to recycle the front surface reflection. One must be cautions though for a cell thinner than $100 nm$ ray-optics begin to loos it grip to the wave description of light. The effectiveness of ray-optics based TIR trapping is thus challenged in less then $100 nm$ a-Si:H layers - an effect not accounted for in our analysis and thus indicated by dashing the respective orange line. This is also why the combination of external trap with internal TIR trapping was not considered in this case. The performance of external-trapping, on the other hand, is not challenged by subwavelength thin layers since its reliance on ray-optics is based on the size of trap, not the cell \cite{Niv2016a}. Note that the $70 nm$ a-Si:H layer with back-reflector and external-trap has the same absorption as $200 nm$ a-Si:H layer with TIR trapping, and by $60 nm$ it outperforms that of the back-reflector only. This is an important point for a-Si:H cells since the light induced degradation is less present in thinner layers, which show the potential of external-trap to revive a material doomed by conventional paradigm of cell construction.
	\paragraph{Perovskites:} After seeing what benefit can arise from an external-trap for existing solar cell materials such as crystalline silicon, or the potential to revive an obsolete material in the form of hydrogenated amorphous silicon, we now turn our attention to what can be considered the future of solar cell material in the form of perovskites. Accordingly, we insert the following parameters to our solar cell model: $1.6 eV$ for the bandgap, $5$ for the relative non-radiative coefficient, $2.75$ for the refractive index, and $9\times10^4 cm^{-1}$ for the absorption coefficient \cite{jeon2015compositional,sha2015efficiency,samiee2014defect,lin2015electro}. Note that these parameter represent what can be considered a typical perovskite since many variants of this material are currently being studied. As in previous cases, all other trap and environmental parameters are left the same. Figure \ref{fig:perovskite-cell} shows the results in this case. Since perovskite has the highest absorption among the materials here studied, it also takes the thinnest layer for maximal absorption, which is $200 nm$ in this case. Here, as in previous cases, the ability to recover the front surface reflection is the main advantage over the back reflector. For thin layers, the external-trap can match the back reflector case with only $~20 nm$ of perovskite. TIR-based trapping was not considered here since no account of it for perovskites has been found which is, as for a-Si:H, not surprising given its high absorption and often subwavelength thickness. The $V_{oc}$ plots shows why perovskite is thought after: The combination of high optical absorption with low rate of non-radiative recombination gives a steady decrease for thinning layer. This in contrast with both c-Si and a-Si:H where more bulk nonradiative recombination gave a peak in $V_{oc}$. Still, there's a benefit in efficiency from using an external trap: The perovskite layer can be made almost 10 time thinner, $20 nm$ with an external-trap relative to $200 nm$ with back-reflector only in our case, while having the same efficiency. This brings forth more the possibility o utilizing more stable perovskites that were so far unfavorable due to their smaller optical absorption \cite{chen2019highly}. The fact that 'only' $19.5 \%$ efficiency arises, in this case, is meaningless since it merely represents the choice of material parameters that we have taken. 
	
	\paragraph{The role of the trap geometry:}
	It is possible that a hemispherical shape is not always the best choice as an external light-trap. For example, if square cell is considered, then the circular base of the hemisphere has to inscribe the cell. The wall area, in this case, increases beyond what is necessary to cover the cell so the wall absorption increases and the effectiveness of the trap reduce. Hence, it is advisable to explore traps of various shapes. Within our flux-balancing approach, the shape of trap is accounted for only by the view-factor $F$, which, in principle, allow us to tackle any form. In practice, however, the view factor, as seen in Eq. (\ref{eq:dV-factor}), gives simple integrable expressions only for highly-symmetrically shaped traps, as discussed in section \ref{sec:empty_trap}. Therefore, despite the reasonable agreement that have emerged between our flux-balance approach and the  measured radiance in a \emph{spherical} trap, as discussed in section \ref{sec:exp res}, adaptation of our formulation to different shapes may prove challenging.
	\par
\begin{figure}[h]
		\centering
		\includegraphics[width=\textwidth]{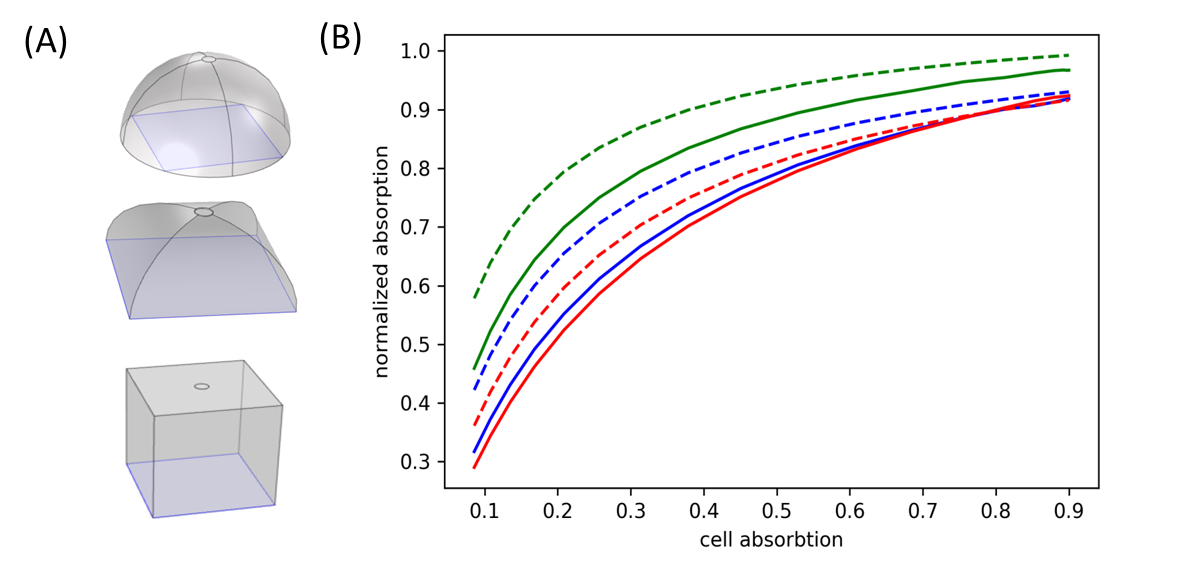}
		\caption{ (A) The different cavity shapes under consideration, from top to bottom: hemispherical, domical, and cube. The cell, in each case is depicted as purple rectangular at the base of the trap. (B) The absorption of a trap equipped cell as a function of the bare cell absorption; green - domical; blue - hemisphere; red - cube. Solid and dashed lines corresponds to the results of the simulation and the flux-balancing formulation of Eq. (\ref{eq:center-mounted-cell-abs}), respectively.}
		\label{fig:comsol}
	\end{figure}

 In order to mitigate the shortcoming in view-factor calculations, we adopted ray-tracing simulation, which has been used to optimize integrating spheres \cite{Prokhorov2003}, solar-cavities, and solar-concentrators \cite{Shuai2008}. We have, therefore, chose a commercial ray-tracing tool to study hemispherical, cubic and domical shaped traps, as seen in Fig. \ref{fig:comsol}(A). By 'domical' we refer to the intersection of two half-cylinders having mutually perpendicular axes, a shape also known as a \emph{cloister vault} in the literature. Hemispherical and cubic light-traps have been studied before with a similar tool \cite{Weinstein2014, Valades-Pelayo2016}; the domical shape is considered since it is, in fact, the square-based equivalent of the hemisphere. Taking $1\,cm^2$ cell, wall-area of $2\,cm^2$ emerges for the domical trap, where $3.7\,cm^2$ and $5\,cm^2$ emerges for the hemispherical and cube shaped traps, respectively. The cell is seen as purple square patch at the bases of different shapes. Since a circular cross-sectional illumination was defined, some of the impingement light inevitably falls on the cavity-wall first, rather then on the cell. The amount of which depends on the trap geometry. In all cases, port-area was kept at $1\,mm^{2}$ and $95\%$ Lambertian wall reflection (at random direction) was considered. 
	\par Figure \ref{fig:comsol}(B) compares the cell absorption from our flux-balance formalism (dashed) to the ray-tracing (solid) result. The plots confirms once more the importance of minimizing the wall area: Even for $95\,\%$ reflection that can be considered high from a practical point of view, the domical shaped trap, which has the least area, appears to be the most effective one. On the contrary, for low to moderate cell-absorption, the cube, which has the  larges wall area, is the inferior trap. This changes once cell-absorption surpasses a value of $0.8$ due to the imperfect coverage of the square cell on the circular base of the hemispherical trap. It is also seen that the flux-balance approach overestimates the absorption. This because the view-factor was evaluated for a small cell at the middle of a big hemispherical trap. Minimizing the wall area caused the cell to reach the side-walls of the trap where the view-factor is, in fact, smaller than that at the center. The fact that the deviation between calculation and simulation is larger for smaller cell-absorptions is because recycled light takes a larger portion of the absorption in this case, hence, the error due to overestimated view-factor is more pronounced in this case. The most encouraging observation is, perhaps, that our somewhat na\"ive flux-balancing approach can reproduce quite faithfully low-symmetry shaped traps.  
	\paragraph{The role of diffused light:}
	Finally, there's the issue of diffused ambient light: It is well know that due to atmospheric scattering and ground reflection, sometimes as much as $30\%$ of the light is present in the form of diffused light. Bare panels make better use this light compared to concentrated ones - a major cause for concentrator photovoltaics (CPV) to fall out of grace albeit being more efficient. In that regard, note that the role of concentrator in above mentioned trapping scheme, as seen Fig. \ref{fig:center-mounted-cell} or \ref{fig:wall-mounted-cell}, is not to concentrate light on the cell, but onto the port of the trap. Since the concentrator role is primarily trapping, the solid angle span by the light as it enters the trap needs only to cover the cell. The external trap, on the other hand, handles light encompassing the entire $2\pi$ solid angle. This means that some of diffused light can be funneled into the trap an eventually be converted to electricity by the cell, albeit at lesser efficiency since the first incidence is with the cavity-wall rather then the cell.
	\par To demonstrate this point let us consider the hemispherical trap from Fig. \ref{fig:dome}, a $1\,(cm^2)$ rectangular cell, a radius of $\sqrt{2}/2\,()cm)$, and a base area of $A=\pi/2\,(cm^2)$. Let us also consider a non-imaging concentrator which is placed atop the entrance port of the trap. We take the input port of the concentrator to be the same area as the base of the trap, namely $A_{in}=\pi/2\,(cm^2)$. The geometric concentration factor of the cell is little less of one in this case. In addition, the port-area, which is also the exist port of the concentrator, is $A_{port}=0.01\,cm^2$. We shall use \'etendu (solid angle to area product) to determine what is the maximal reception area of this concentrator. Conservation of \'etendue implies: $A_{in}\Omega_{in}=A_{out}\Omega_{out}$. Taking $A_{out}=A_{port}=0.01$, $\Omega_{out}=2\pi\,(srad)$, and $A_{out}=pi/2\,(cm^2)$, we find $\Omega_{in}=2\pi\,A_{out}/A_{in}=0.04\,(srad)$. In terms of angles this corresponds to $6.5\,(deg)$ with respect to the concentrator entrance-port normal. Therefore, the trap can capture a fan  of rays spanning $13\,(deg)$ allowing at least some of the diffused light to be transferred to work. One should be mindful, however, that the trap needs an exquisite wall reflection to make good use of this diffused light since only a small portion of it impinges the cell at first instance.
	
\section{Summary and conclusions}
	External light-traps have the ability to decouple the optics from the electronic aspects of solar power generation. As such, they enable the design of solar cells that excel in charge separation and transport, by being thinner or having larger front grid electrodes, on the expense of their ability to absorb light. The function of the external trap, in this case, is to re-direct light reflected at first instance with the cell back for secondary absorption. The combined cell-trap system establishes an optical feedback-loop that produces more current at a higher voltage and therefore is more efficient than what a cell from the same material can achieve by its own.
	\par Despite the fact that external light-traps have been discussed for quite a while, no comprehensive analysis of their potential has yet emerged. We hereby attempt to do so for a Lambertian inner wall reflecting trap. This opposes the more common specular inner wall reflection type \cite{luque1991confinement,Braun2013,Weinstein2015,VanDijk2016a}. Considering perfect ray direction randomization inside the trap, as Lambertian reflection entitles, we use statistical ray-optics flux-balancing approach to derive the relation between the entering power and the internal radiance of the trap. We then use this expression to derive the power absorbed by the cell. The trap geometry and that of the cell as well as the cell's position, enter the formalism through a view-factor term borrowed from non-imaging optics. We then measured the power capture by a detector of a spectroscopic integrating sphere with different port areas and sizes of absorbing foils inserted inside. The results are in good agreement with the predictions of our flux-balancing formulation.
	\par The analysis points to the fact that it is better to put the cell at the center of the trap and to minimize the trap wall-area, which is the major loss mechanism in this case. Based on this conclusions we examine what would be the benefit from using a realistic trap for cells made from three kinds of semiconductors: crystalline-silicon, amorphous hydrogenated-silicon, and perovskite. The solar cell is modeled, in this case, according to the Shockley and Queisser detailed-balance approach, including the effect of Shockley-Read-Hall type non-radiative recombinations \cite{Queisser1961}. The results for all three is that active layer can become much thinner while the efficiency slightly increase. The analysis also shows that the front surface reflection of the cell can be made substantially bigger without sacrificing the absorption of the combined cell-trap arrangement. The effect that wider metal grid electrodes has on the efficiency, a major cause of front surface reflection, was not considered. Also, recapturing the cell's own bandgap emission, the so called \emph{photon recycling}, which was the original motivation for one of the earliest experimental demonstration of an external light-trap \cite{Braun2013}, is yet another effect that was neglected. Therefore, the actual benefit from an external trap may be larger than what is shown here. This conclusion is particularly true for silicon, since Auger non-radiative recombination was also not considered.
	\par We have complemented our analytical treatment with ray-optics simulation of trap which is less symmetrical than the hemisphere. Results of which shows that minimizing the wall-area is more important than the shape of the trap in this case - a hallmark of Lambertian trapping. Results also show that the analytical formulation holds true even if for these less symmetric traps.
	\par External trapping is always supplemented with a solar concentrator, which raises the question as to the ability to use diffused ambient light. We argue that since the primer objective of the concentrator, in this case, is trapping and not, necessarily, concentrating the light onto the cell, at least some of the diffused-light can make its way into the trap. Diffused ambient light, is therefore, less of a problem in this case with respect to what it is in concentrated photovoltaics. As an example, a realistic trap capturing $13\degree$ fan of rays is described.  
	\par Based on our analysis, which was validated experimentally simulation-wise, we conclude that an external light traps, if properly designed, may allow much less absorptive cells operate at higher efficiently, although not by much. This is exceptionally true for crystalline-silicon if external and internal TIR based trapping are combined. Therefore, Lambertian external light-traps are beneficial mainly for cases where thinning of the active layer of the cell or making wider front grid electrodes is desired.
	
\begin{appendices}
	\setcounter{equation}{0}
	\setcounter{figure}{0}
	\setcounter{table}{0}
	\renewcommand{\theequation}{\thesection.\arabic{equation}}
	\renewcommand{\thefigure}{\thesection.\arabic{figure}}
	\renewcommand{\thetable}{\thesection.\arabic{table}}    
	\section{Radiance on a wall mounted cell with a center absorber and its fitting to the measured data}\label{app: model and fitting}
\begin{figure}
			\centering
			\includegraphics[width=.8\textwidth]{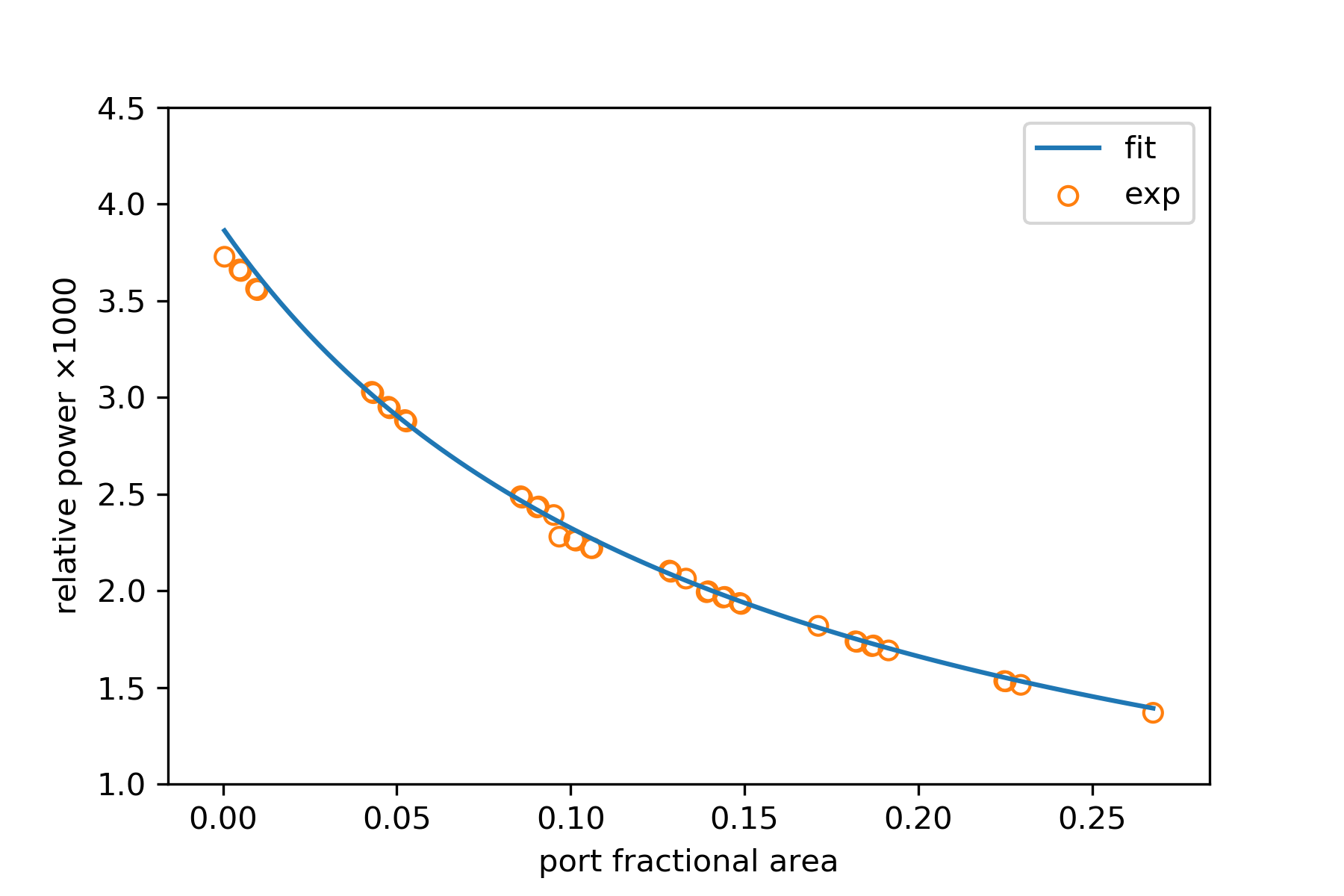}
			\caption{\small{Relative measured power and the fitting for an empty modeled-trap. Port-area is given as the fraction of the total area, port+wall.}}
			\label{fig:empty-reflector-fit}
		\end{figure}

		Experimental results were obtained by measuring the power inside an integrating sphere acting as an external light-trap. Three different scenarios were progressively considered: (1) Empty trap; (2) Trap with only the sample holder inside; (3) Trap loaded with the different samples. Measured powers from each scenario was normalized by the registered power at the input port. According to the formalism from section \ref{sec:empty_trap}, the relative power at the detector for the empty trap is:
		\begin{equation}\label{eq:empty-cav}
			\frac{P_d}{P_{in}}=\frac{\alpha_d A_d (1-\alpha_w)}{\alpha_p A_p + \alpha_w A_w + \alpha_d A_d}
		\end{equation}
		The fitting of this case is shown in Fig. \ref{fig:empty-reflector-fit}, which yielded values for $\alpha_{d}$, $\alpha_{w}$, and $\alpha_{p}$.
		\par
\begin{figure}	
			\centering
			\includegraphics[width=.8\textwidth]{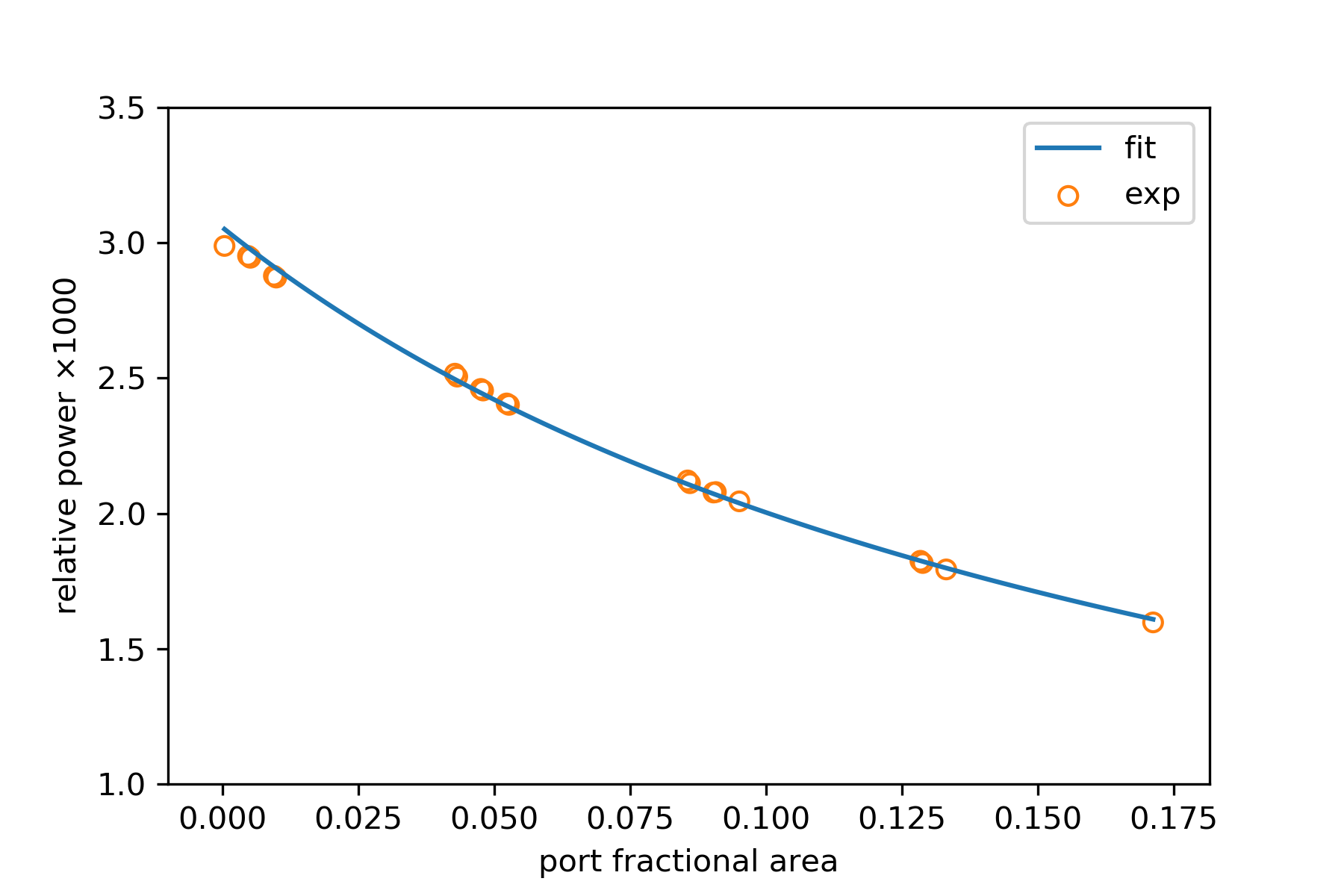}
			\caption{\small{Relative measured power and the fitting for the modeled trap and the sample holder. Port-area is given as the fraction of the total area, port+wall.}}
			\label{fig:reflector-holder-fit}
		\end{figure}

 Next, the sample holder was mounted, which introduced additional absorption, and so the normalized power becomes:
		\begin{equation}\label{holder-cav}
			\frac{P_d}{P_{in}}=\frac{\alpha_d A_d (1-\alpha_w)}{\alpha_p A_p + \alpha_w A_w + \alpha_d A_d + \alpha_h A_h}
		\end{equation}
		The fitting of this case is shows in Fig. \ref{fig:reflector-holder-fit}, which yielded the additional normalized absorption coefficient $\alpha_h A_h$.
		\par
\begin{figure}
			\centering
			\includegraphics[width=.8\textwidth]{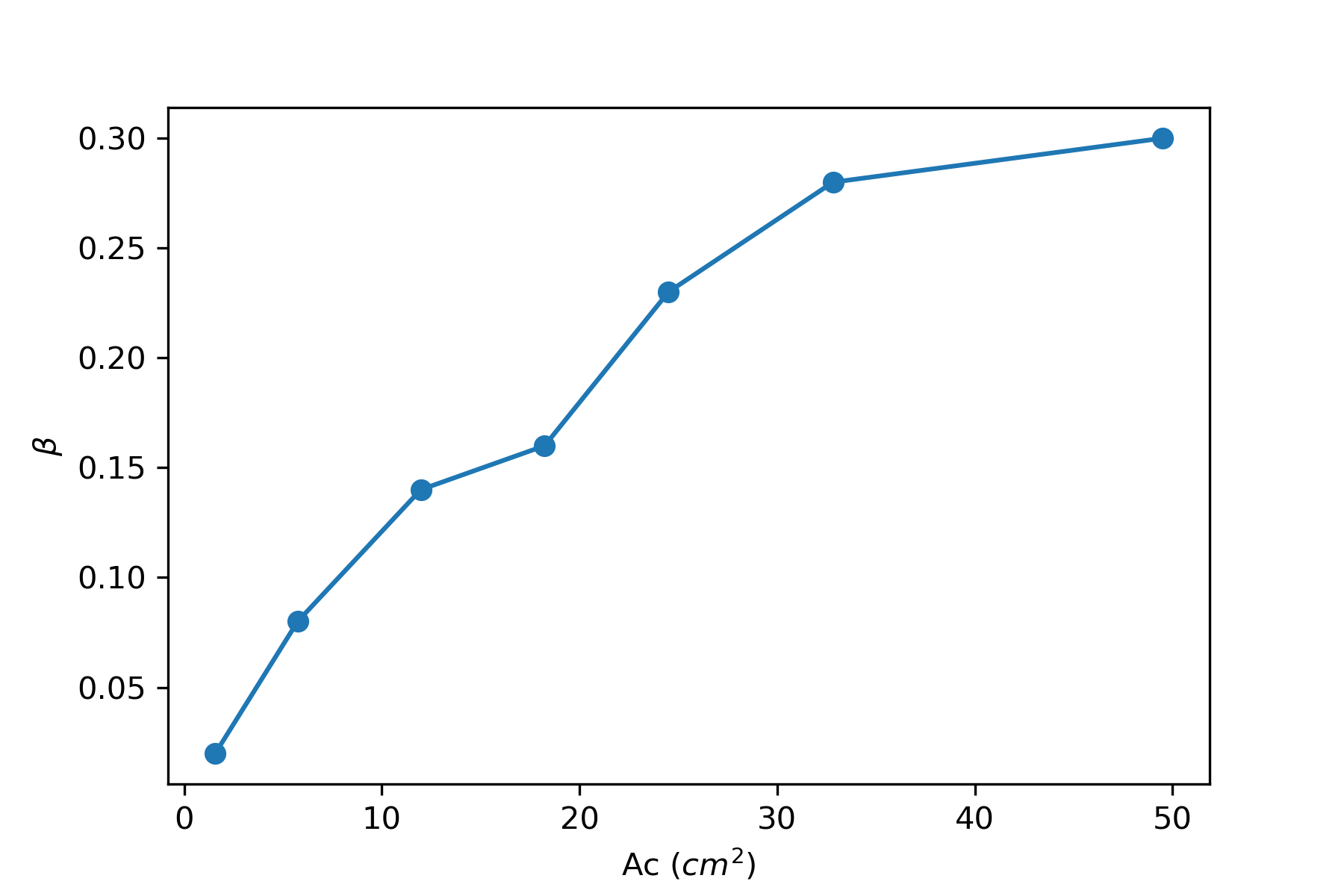}
			\caption{\small{Values of the direct illumination factor $\beta$.}}
			\label{fig:beta}
		\end{figure}

\begin{figure}
			\centering
			\includegraphics[width=.8\textwidth]{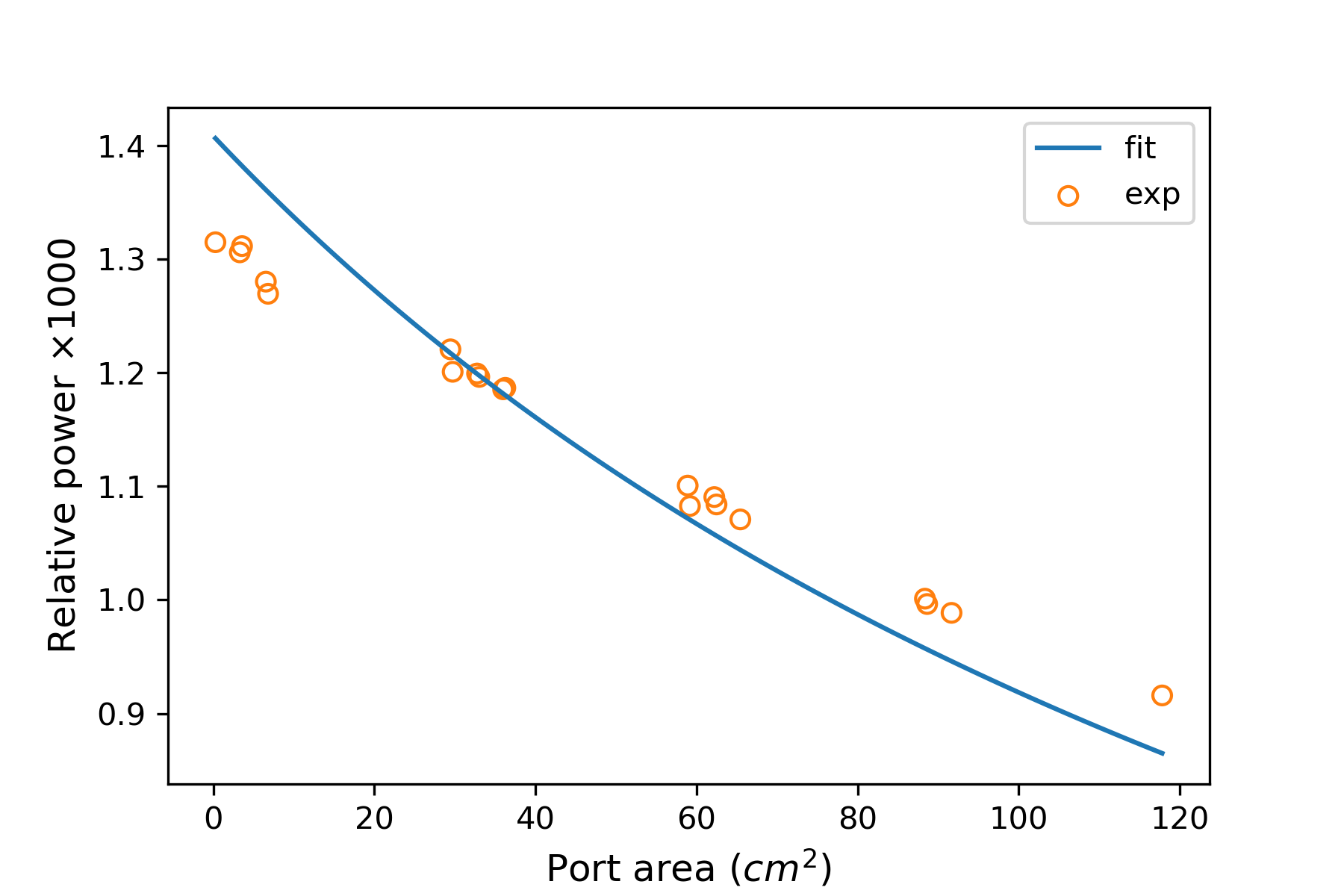}
			\caption{\small{Relative power for the $33\,cm^2$ aluminum foil tilted $10\degree$ to the beam axis.}}
			\label{fig:tilt}
		\end{figure}

 Finally, different samples, in the form of dark-anodized aluminum foils, were mounted on the holder and their relative power versus port-area was measured. The foils were mounted parallel to the incoming beam and to the detector axis to reduce direct illumination effects and so maximize the buildup of diffused radiation power inside the trap. Normalized detector absorption, in this case, is given by:
		\begin{equation}\label{eq:cavitiy-targ}
			\frac{P_d}{P_{in}}=\frac{\alpha_d A_d ((1-\alpha_{c})\beta + (1-\alpha_w)(1-\beta)}{\alpha_p A_p + \alpha_w A_w + \alpha_d A_d + \alpha_h A_h + \frac{3}{2} \alpha_{c} A_c}
		\end{equation}
		Two more fitting parameters are now added: $\alpha_{c}$ for the sample normalized absorptivity and $\beta$ for the direct illumination factor. The nominal sample area considers the double-sided absorption. View factor of $3/2$ is taken for the center-mounted sample, in agreement with section \ref{sec:empty_trap}. The result of this fit is shown in Fig. \ref{fig:theory-exp}, with the $\beta$ values gotten for each foil area shown in Fig. \ref{fig:beta}; monotonic increase is only expected given the progressively larger foil area. Finally the relative power of the $33\,cm^2$ foil tilted by $10\degree$ are shown in Fig. \ref{fig:tilt}. Fitting was obtained, in this case, by taking $\beta = 0.34$ instead of the former $0.28$. From purely geometrical perspective, the tilted value should be $\beta_{tilt} = (1+\sin(10\degree)\times0.28)= 0.328$, in agreement with the fitted value.
		
		\par The trap parameters that were obtained with this progressive fit-procedure are summarized in table \ref{table:fitparam}. All are reasonable but the detector absorptivity $\alpha_d$, which is larger than one, a discrepancy attributed to the inner mounted shading screen that altered the detector view-factor with respect to a direct exposure to the incoming beam. The factor $\alpha_h A_h=24$ designates area-absorptivity product of the sample holder. Given the different port areas, the corresponding factor for the cavity wall, namely $\alpha_{w}A_{w}$, changes from $74$ and up to $89\,cm^2$. The holder absorption is, therefore,  considerable but not a dominant. The good fit proves that the proposed formalism can predict the performance of an external diffused light-trap.
		\begin{table}[h!]
			\centering
			\begin{tabular}{ |c|c|c|c|c| } 
				\hline
				$\alpha_d$ & $\alpha_w$ & $\alpha_p$ & $\alpha_h A_h\,(cm^2)$ & $\alpha_{c}$ \\
				\hline
				2.4 & 0.13 & 1 & 24 & 1 \\
				\hline
			\end{tabular}
			\caption{\small{Fitting parameters.}}
			\label{table:fitparam}
		\end{table}

	\setcounter{equation}{0}
	\setcounter{figure}{0}
	\setcounter{table}{0}
	
	\section{Analytic model for the open circuit voltage, short circuit current, and efficiency of a solar cell with internal trapping}\label{app:cell-model}
		We hereby present an analytic model for a solar cell which is based on Shockley and Queisser detailed-balance analysis from Ref. \cite{Queisser1961}. The cell's characteristic current-voltage relation, in this case, is given by:
		\begin{equation}\label{eq:I-V}
		\Omega_s\,A(aW)\,\frac{{T_s}^3}{h^3 c^2}\,R\left(\frac{E_g}{T_s}\right) = \pi\,\frac{{T_c}^3}{h^3\,c^2}\,R\left(\frac{E_g}{T_c}\right)\left(1+r_{NR}W\right)(e^{V/T_c}-1) + \frac{1}{q_e}I,
		\end{equation}
		On the left there is the generation rate of electron-hole pairs due to absorption of sunlight, which is modeled here as a thermal source with temperature of $T_s=0.51 (wV)$, $(\approx 5900K)$. The generation rate is proportional to the solid angle of the sun $\Omega_s=6.85\times10^{-5} (sr)$ and the normalized absorption which is a function of the cell's absorptivity $a (cm^{-1})$ times its thickness $W (cm)$. The particular form of the function $A(aW)$ depends on the trapping mechanism being used. On the right are radiative and non-radiative recombination rates at the cell's temperature $T_c=0.025 (eV)$, $(\approx290K)$, with $V$ and $I$ denoting the voltage and the current drawn from the cell, respectively. The non-radiative rate coefficient is $r_{NR} (cm^{-1})$, is the Shockley-Read-Hall kind. The cells emission is considered flat-plane Lambertian and hence proportional to $\pi (sr)$. Finally, $h$, $c$, and $q_e$ are Plank's constant, the speed of light, and the elementary charge, respectively, and the unitless black-body emission-rate is:		
		\begin{equation*}
		R(x)=\int_{x}^{\infty}\!\frac{y^2}{e^{y}-1}\,dy.
		\end{equation*}
		\par The power from the cell is, in this case:
		\begin{equation}\label{eq:power}
		P=FF(V_{oc}/T_c)I_{sc}V_{oc},
		\end{equation}
		where the short-circuit current is gotten form \ref{eq:I-V} by taking $V=0$:
		\begin{equation}\label{eq:Isc}
		I_{sc} = \Omega_s\,q_e\, A(aW)\frac{{T_s}^3}{h^3\,c^2} R\left(\frac{E_g}{T_s}\right),
		\end{equation}
		and the open circuit voltage by taking $I=0$:
		\begin{equation}\label{eq:Voc}
		V_{oc} = T_c\,\ln\left(\frac{\Omega_s\, A(aW)\,{T_s}^3 R\left(\frac{E_g}{T_s}\right)}{\pi\,{T_c}^3\,R\left(\frac{E_g}{T_c}\right)(1+r_{NR}W)}+1\right).
		\end{equation}
		Finally, the fill-factor given by the phenomenological expression \cite{green2006third}:
		\begin{equation}\label{eq:FF}
		FF(x)=(x-log(x+0.72))/(x+1).
		\end{equation}
		\par The normalized absorption $A(aw)$ is determined by the light-trapping mechanism that the cell possess. Several light trapping mechanisms are considered: The first is a perfect back reflector. In this case the trivial double pass absorption emerges:
		\begin{equation*}
			A_{DP}(aW)=(1-R_{ext})(1-\exp(-2\,a\,W)),
		\end{equation*}
		where $R_{ext}$ is the external reflection from the cell's front surface.
		Next, there's the absorption that emerges from internal TIR trapping, modeled here after Gee as \cite{Gee1988}:
		\begin{equation*}
			A_{Lam}=\frac{(1-R_{ext})(1-T_b)(1-T_bR_{bsr})}{1-T_b^2R_{bsr}R_{int}},
		\end{equation*} 
		with $R_{bsr}$ the back surface reflection, $R_{int}=1-(1-R_{ext}/n^2)$ the internal front surface reflection, and the normalized transmission through the bulk of the cell:
		\begin{equation*}
		T_b=2\int_0^1y\exp(-a\,W/y)\,dy
		\end{equation*}
		Finally, there's our model for the wall or center mounted cell inside an external light-trap. Since these are irrespective of the internal light-trapping being used, we simply need to insert the above mentioned $A_{DP}$ and $A_{Lam}$ into Eq. (\ref{eq:wall-mounted-cell-abs}) or (\ref{eq:center-mounted-cell-abs}) to find the power-absorption for each case.
\end{appendices}

\newpage
\bibliographystyle{unsrt}
\bibliography{cavity_bib}

\end{document}